\documentclass[12pt]{article}
 \usepackage[dvips,final]{graphicx}
  \usepackage{amsmath}
   \usepackage{amssymb}
    \usepackage{pifont}

\begin{document}

\begin{center}
{\Large Accessing the pion distribution amplitude
       through the CLEO and E791 data}\\ [0.5cm]

\textbf{Alexander~P.~Bakulev}\footnote{E-mail: bakulev@thsun1.jinr.ru},\ \
\textbf{S.~V.~Mikhailov}\footnote{E-mail: mikhs@thsun1.jinr.ru}\\[0.5cm]

\textit{Joint Institute for Nuclear Research,
Bogoliubov Lab. of Theoretical Physics,\\
141920, Moscow Region, Dubna, Russia}\\ [0.5cm]

\textbf{N.~G.~Stefanis}\footnote{E-mail:
                                 stefanis@tp2.ruhr-uni-bochum.de}\\[0.5cm]

\textit{Institut f\"ur Theoretische Physik II,
       Ruhr-Universit\"at Bochum,\\
      I-44780 Bochum, Germany}\\ [0.5cm]

\end{center}

\begin{abstract}
Using QCD perturbation theory in NLO and light-cone QCD sum rules,
we extract from the CLEO experimental data on the
$F^{\gamma^*\gamma\pi}\left(Q^{2}\right)$ transition form factor
constraints on the Gegenbauer coefficients $a_2$ and $a_4$,
as well as on the inverse moment $\langle{x^{-1}}\rangle_{\pi}$ 
of the pion distribution amplitude.
We show that both the asymptotic and the Chernyak--Zhitnitsky
pion distribution amplitudes are excluded
at the $3\sigma$- and $4\sigma$-level, respectively,
while the data confirms the
end-point suppressed shape of the pion DA we previously obtained
with QCD sum rules and nonlocal condensates. 
These findings are also supported by the data 
of the Fermilab E791 experiment on diffractive dijet production.
\end{abstract}
\vspace {2mm}



\section{Introduction}
Perturbative QCD describes the short-distance interactions 
of quarks and gluons 
and can be applied to the description of hadronic reactions 
on account of factorization theorems. 
More precisely, 
one can calculate systematically perturbative kernels
and associated anomalous dimensions that govern the evolution 
of hadron distribution amplitudes (DAs). 
These DAs parameterize hadronic matrix elements 
of quark-gluon currents and have to be determined 
by nonperturbative methods or extracted from experimental data. 
Recently, Schmedding and Yakovlev \cite{SchmYa99} 
have presented an analysis, 
based on light-cone QCD sum rules (LCSR) 
proposed earlier by Khodjamirian \cite{Kho99}
and taking into account $O(\alpha_s)$-corrections, 
of the high-precision CLEO experimental data \cite{CLEO98} 
that allows to extract 
quite restrictive constraints 
on the first two Gegenbauer coefficients $a_2$ and $a_4$ 
which control the $x$-dependence
of the pion distribution amplitude ($\pi$DA). 
This sort of analysis 
was further extended and refined by us in \cite{BMS01,BMS02} 
with the aim 
to take more properly into account NLO evolution effects 
of the $\pi$DA, 
to treat threshold effects 
of the effective strong coupling, 
and to estimate more carefully contributions resulting 
from (unknown) higher-twist effects. 
In addition, we derived directly from the CLEO data estimates 
for the inverse moment of the $\pi$DA, 
which is compatible with that obtained 
from an \textit{independent} QCD sum rule, 
referring in both cases to the same low-momentum scale
of the order of 1~GeV$^2$.

The results of our analysis, presented here,
lead to the conclusion
that the Chernyak--Zhitnitsky model \cite{CZ84}
for the $\pi$DA in the plane $(a_2,\, a_4)$
is outside the $4\sigma$-level,
while the asymptotic DA,
is excluded at the $3\sigma$ level.
In fact, the data seem to prefer end-point-suppressed DAs
as those we have previously determined using QCD sum rules
with nonlocal condensates \cite{BMS01}.
These conclusions are further supported
by contrasting the above mentioned $\pi$DAs
with the E791 dijet data \cite{E79102}
following the convolution approach of Braun et al. \cite{BISS02}.
Moreover, it was found \cite{BMS02} that the CLEO data are sensitive
to the value of the average vacuum quark virtuality,
limiting its value close to $\lambda_q^2=0.4$~GeV$^2$.

\section{What is the pion distribution amplitude $\varphi_{\pi}(x,\mu^2)$?}

\begin{figure}[hb]
 \centering{\includegraphics[width=0.63\textwidth]{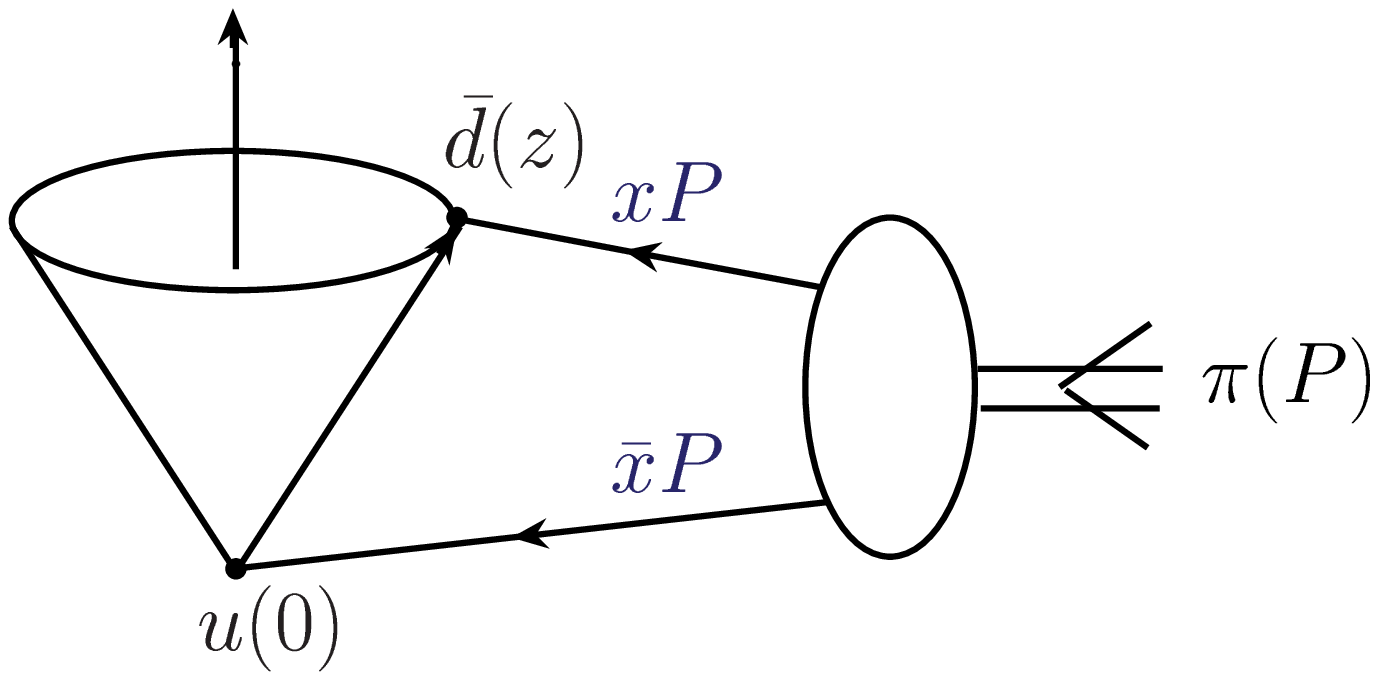}}
  \caption{$\varphi_\pi(x;\mu^2)$ -- light-cone amplitude for the
   transition $\pi \rightarrow u + d$.}
   \label{fig-1}
\end{figure}

The $\pi$DA is a central object in the deeper understanding of the
pion microscopic structure in terms of quark and gluon degrees of
freedom within QCD. This amplitude is defined by the matrix
element of a nonlocal axial current on the light cone:
\vspace*{-3mm}
 \begin{eqnarray}
  \langle 0|\bar d(z)\gamma_{\mu}\gamma_5{E(z,0)}u(0)|\pi(P)\rangle\Big|_{z^2=0}
  &=& i f_{\pi}P_{\mu}
   \int_{0}^{1} dx\ e^{ix(zP)}\
     {\varphi_{\pi}^\text{Tw-2}(x,\mu^2)}\; ,
    \end{eqnarray}
where gauge-invariance is ensured due to the Fock--Schwinger
string $E(z,0) = {\cal P}e^{i g \int_0^z A_\mu(\tau) d\tau^\mu}$
and ${\varphi_{\pi}^\text{Tw-2}(x,\mu^2)}$ is symmetric with
respect to $x\leftrightarrow \bar{x}$ ($\bar{x}\equiv 1-x$) and is
normalized to unity, whereas $\mu^2$ denotes the normalization
scale. Fig.~\ref{fig-1} visualizes the light-cone structure of the
$\pi$DA.
There are also 6 pion DAs at twist-4 level, 
four of them contributing to the $\gamma^*\gamma\to\pi$-transition 
as a twist-4 correction, 
whose value is parameterized by the scale 
$\delta_\text{Tw-4}^2\approx0.19\text{ GeV}^2$. 
In what follows we will speak mainly of the twist-2 $\pi$DA 
and for the sake of brevity we will omit 
its superscript $^\text{Tw-2}$ 
referring to it simply as $\varphi_{\pi}(x;\mu^2)$.

Due to vector current conservation, 
the solution of the ERBL evolution equation~\cite{ER80,BL79} 
(in LO approximation) in the asymptotic limit 
is ${\varphi_\pi(x;\mu^2\to\infty)} = \varphi^\text{As}(x) = 6x(1-x)$. 
A particularly convenient way to represent 
the $\pi$DA is to use its 1-loop eigenfunctions, viz.,
the Gegenbauer polynomials~\cite{ER80}:
\begin{equation}\varphi_\pi(x;\mu^2) = \varphi^\text{As}(x)
    \Bigl[ 1 + {a_2(\mu^2)} {C^{3/2}_2(\xi)}
           + {a_4(\mu^2)} {C^{3/2}_4(\xi)}
           + \text{...} \Bigr]_{\xi\equiv 2x-1}
\end{equation}
with ${C^{3/2}_n(\xi)}$ being the Gegenbauer polynomials 
and the ellipsis denoting still higher-order eigenfunctions 
than displayed. 
In this representation all the dependence of
$\varphi_\pi(x;\mu^2)$ on $\mu^2$ 
is concentrated in the coefficients $a_{n}(\mu^2)$ 
due to the fact that the 1-loop evolution kernel 
has a factorized structure
$V_\text{1-loop}(x,x';\alpha_s) = [\alpha_s/(4\pi)]V_0(x,x')$.
In the NLO approximation the eigenfunctions 
of the evolution kernel inevitably depend 
on $\alpha_s$ and therefore on $\mu^2$. 
Note that because of the symmetry in
$x\leftrightarrow\bar{x}$, 
only even Gegenbauer polynomials contribute.

The high precision of the CLEO data provides the possibility
to extract these important theoretical parameters
($a_2$ and $a_4$)
directly from experiment.
But before we turn to this subject,
let us first give some brief exposition of the theoretical method
to determine the $\pi$DA within the QCD sum-rule approach.

\section{QCD sum rules with nonlocal condensates}
To model the nonlocality of the QCD vacuum, we assume
 $\displaystyle \langle{\bar{q}(0)q(z)}\rangle
 = \langle{\bar{q}(0)q(0)}\rangle e^{-|z^2|\lambda_q^2/8}$,
and similar expressions for other nonlocal condensates (NLCs),
where a single scale parameter 
$\displaystyle \lambda_q^2 = \langle{k^2}\rangle$
was introduced in order to characterize the average momentum of quarks
in the QCD vacuum~\cite{MR89}:
  $$\lambda_q^2 = \left\{
   \begin{array}{ll}
   0.4\pm0.1~\text{GeV}^2 & {\mbox{from QCD SRs~\cite{BI82}}};\\
   0.5\pm0.05~\text{GeV}^2 & {\mbox{from QCD SRs~\cite{piv91}}};\\
   \approx 0.4-0.5~\text{GeV}^2 & {\mbox{from lattice QCD~\cite{DDM99,BM02}}}.
   \end{array} \right.$$
The correlation length $\lambda_q^{-1}\simeq0.3$ Fm $\sim
\rho$-meson size represents the width of the NLC at small
distances. Let us mention that for very large distances ($z\gg 1$
Fm~\cite{BM02}) one may assume another form of the condensate,
given by~\cite{BM95}
 $\displaystyle \langle{\bar{q}(0)q(z)}\rangle \sim %
  \langle{\bar{q}(0)q(0)}\rangle e^{-|z|\Lambda}$ at $|z|\gg 1\text{ Fm}$
(with $\Lambda\simeq450\text{ MeV}$).
This behavior is of no importance in the problem under investigation.

In \cite{BMS01} we have determined all coefficients up to order $n=10$
using QCD sum rules with nonlocal condensates.
It turned out that all coefficients beyond $n=4$
are very small
so that for practical purposes it suffices to model the $\pi$DA
using only $a_2$ and $a_4$.
So, the NLC QCD sum rules produces a whole bunch
of self-consistent 2-parameter model DAs (see Fig.~2a.)
at $\mu^2\simeq 1~\text{GeV}^2$:
\begin{equation}
{\varphi_\pi(x)} = \varphi^{\text{as}}(x)
   \left[1 + {a_2} C^{3/2}_2(2x-1)
           + {a_4} C^{3/2}_4(2x-1) \right]
\end{equation}
with the best-fit model (bold-faced on Fig.~2a) defined by the parameters
\begin{eqnarray}
 a_2^\textbf{b.f.} & = & + 0.188\,;\qquad a_4^\textbf{b.f.}\ =\ -
 0.130\;.
\end{eqnarray}
The admissible regions for the parameters $a_2$, $a_4$ of the
$\pi$DA are presented in Fig.~2b as shaded slanted rectangles and
are shown for different values of $\lambda_q^2$.
\begin{figure}[h]
 $$\includegraphics[width=0.49\textwidth]{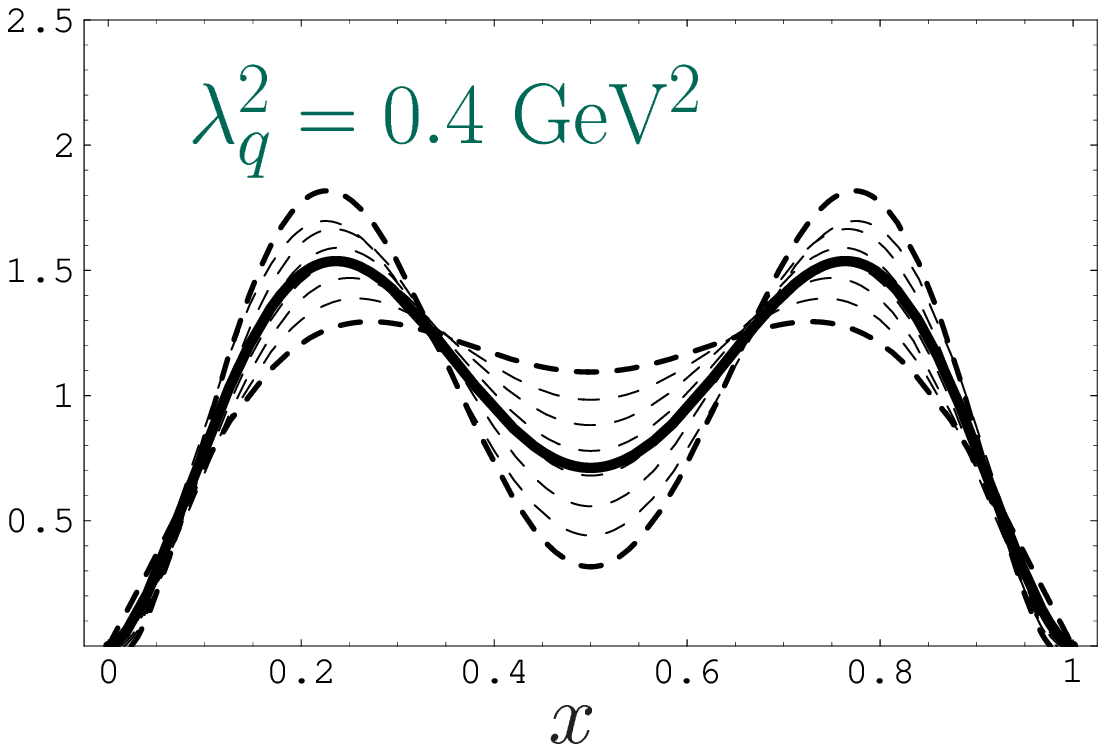}~~~
   \includegraphics[width=0.49\textwidth]{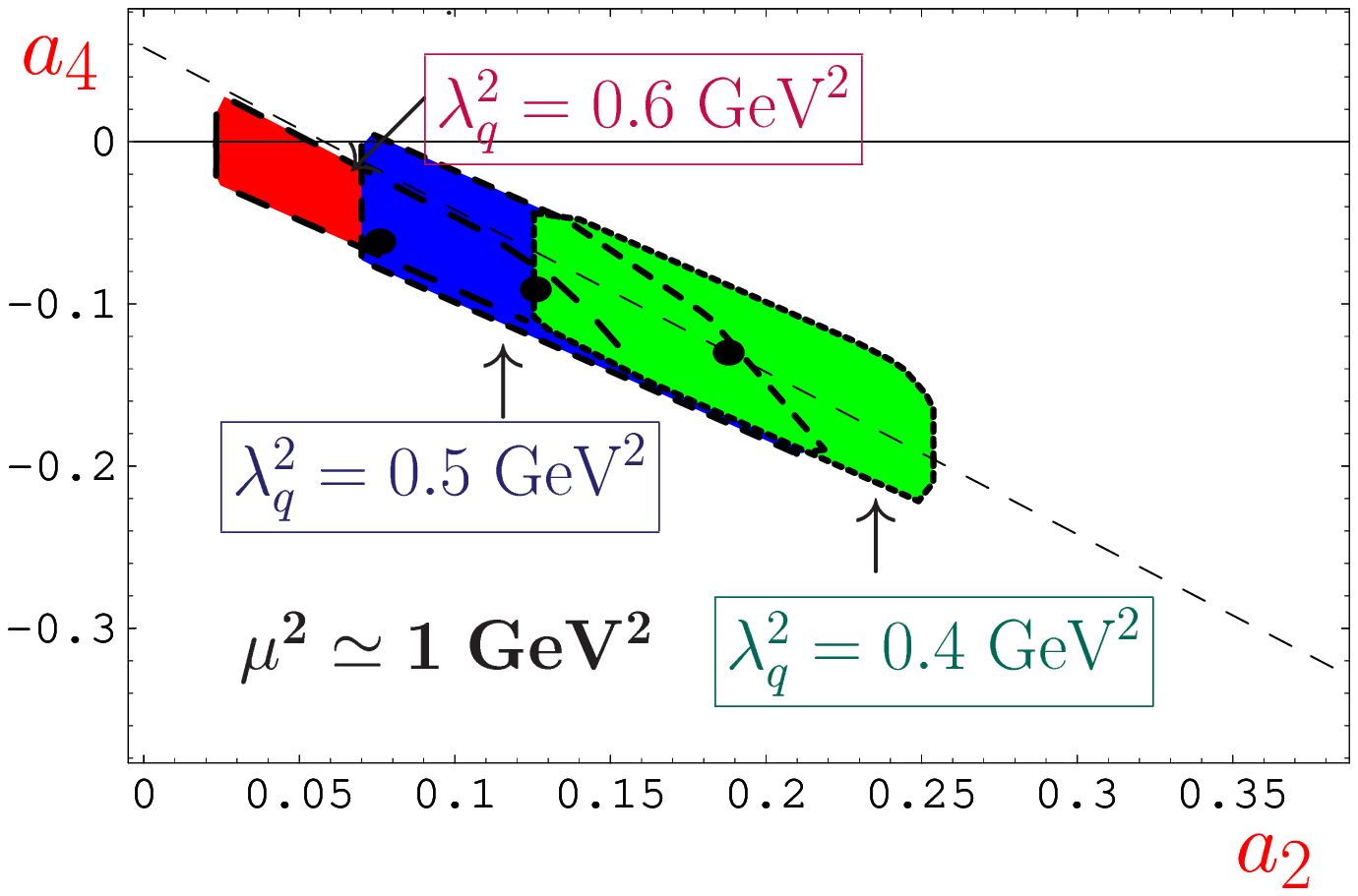}$$
 \caption{\textbf{Left (a):} Self-consistent 2-parameter bunch of admissible $\pi$DAs.
          \textbf{Right (b):} Admissible regions for the parameters $a_2$ and $a_4$ 
          of the $\pi$DA.}
\end{figure}
Fig.~2a demonstrates the most striking feature of our type of
$\pi$DAs: their end-points (i.~e., $x\to 0$ and $x\to 1$) are
strongly suppressed, the suppression being controlled by the quark
vacuum virtuality $\lambda_q^2$. Both the asymptotic and the CZ
$\pi$DAs are \emph{not} end-point suppressed, as we have
quantitatively shown in~\cite{BMS01}. Our models demonstrate by a
precedent that the common statement \textsl{``two-humped $\pi$DAs
are end-point concentrated''} is wrong.

\section{$\gamma^*\gamma\to\pi$: Why Light-Cone Sum Rules?}%
For {$Q^2\gg m_\rho^2,\ \ q^2\ll m_\rho^2$}, pQCD factorization
does not help because it is valid only for leading twist and
therefore higher twists become important~\cite{RR96}.
\begin{figure}[hb]
 $$\includegraphics[width=0.49\textwidth]{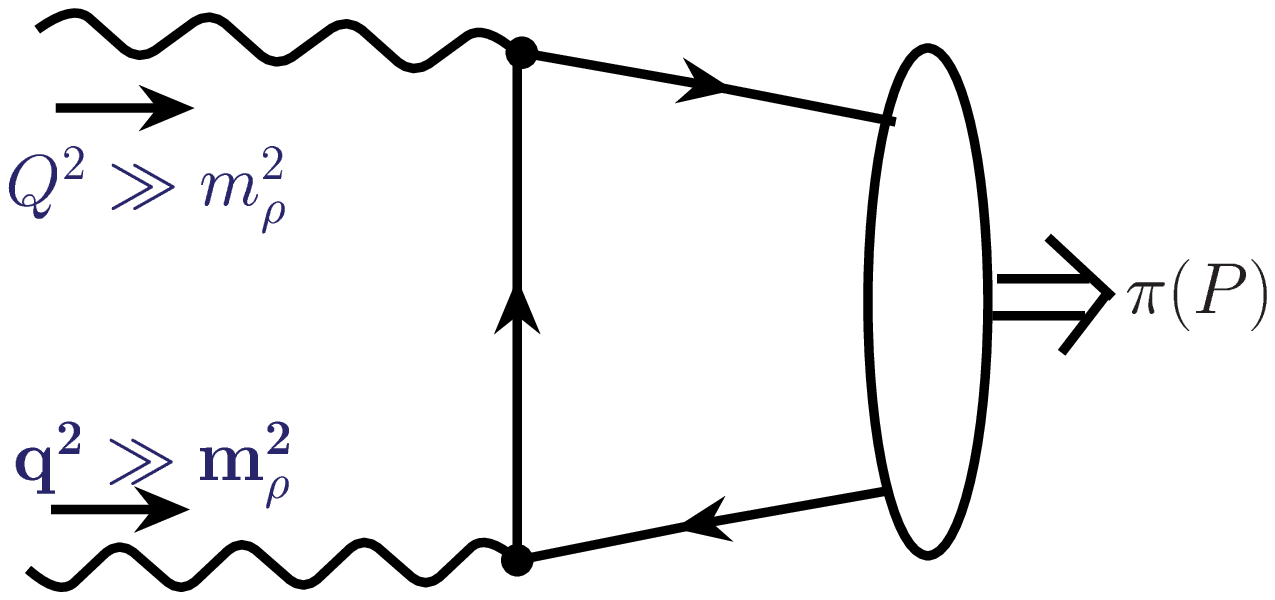}~~
   \includegraphics[width=0.49\textwidth]{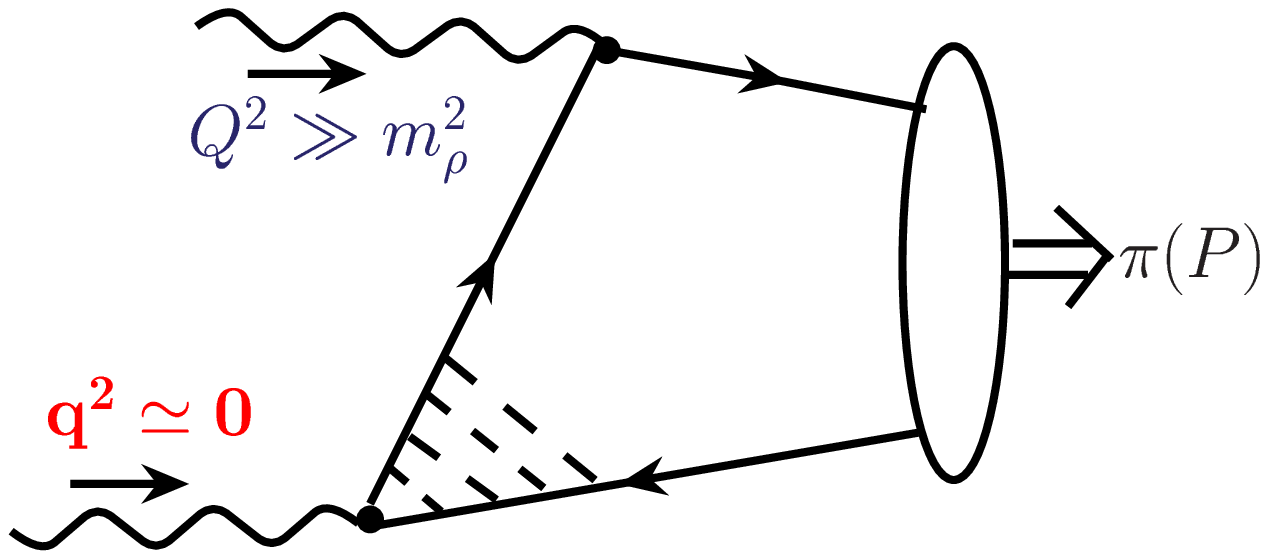}
   $$
 \caption{Left part demonstrates the regime when pQCD description is valid;
          right part makes explicit why LCSR should be applied.}
\end{figure}
The reason for this failure can be understood by recalling that if
$q^2\to 0$, one needs to take into account the interaction of a
real photon at long distances of the order of $O(1/\sqrt{q^2})$,
as the following Fig.~3 illustrates.
To account for long-distance effects in a perturbative QCD treatment,
one needs to introduce the light-cone DA of a real photon.

To this end, Khodjamirian~\cite{Kho99} has shown
that light-cone QCD sum rules (LCSR) effectively account
for the long-distance effects of a real photon
by using quark-hadron duality in the vector channel
and an appropriate dispersion relation in $q^2$; namely,
\begin{eqnarray}
 \label{eq:LCSR}
F_{\gamma\gamma^*\pi}(Q^2,{q^2})
  = \frac{1}{\pi}\int_{0}^{s_0}
   \frac{\rho(Q^2,s)}{m_\rho^2+{q^2}}
     \exp\left[\frac{m_\rho^2-s}{M^2}\right]ds
 + \frac{1}{\pi}\int_{s_0}^{\infty}
    \frac{\rho(Q^2,s)}{s+{q^2}}ds\,,
\end{eqnarray}
where $s_0\simeq1.5\text{ GeV}^2$ is an effective threshold in the
vector channel and the Borel parameter $M^2$ takes values in the
range $0.5-0.9$ GeV$^2$. Then, the real photon limit ($q^2\to 0$)
becomes safely accessible. Here
$\rho(Q^2,s)=\textbf{Im}F_{\gamma^*\gamma^*\pi}^\text{PT}(Q^2,-s)$
includes contributions from both the leading twist $\pi$DA as well
as the twist-4 one. The latter is characterized by the twist-4
scale parameter $\delta^2_\text{Tw-4}$. This theoretical ground
was extended by Schmedding\&Yakovlev (SY) to the NLO
accuracy~\cite{SchmYa99}.

\section{Results from nonlocal QCD sum rules vs CLEO constraints}
In~\cite{BMS02} we improved the SY analysis based on LCSR
(\ref{eq:LCSR}) by taking into account ERBL NLO evolution for the
$\pi$DA and the exact NLO running of $\alpha_s(Q^2)$. The
established relation $\delta^2_\text{Tw-4}\approx\lambda_q^2/2$
has been also involved in the analysis. As Fig.~4a shows, we
obtained reasonable agreement with established in this approach
constraints just for the value of $\lambda_q^2=0.4$ GeV$^2$.
\vspace*{-4mm}

\begin{figure}[hb]%
 $$\includegraphics[width=0.49\textwidth]{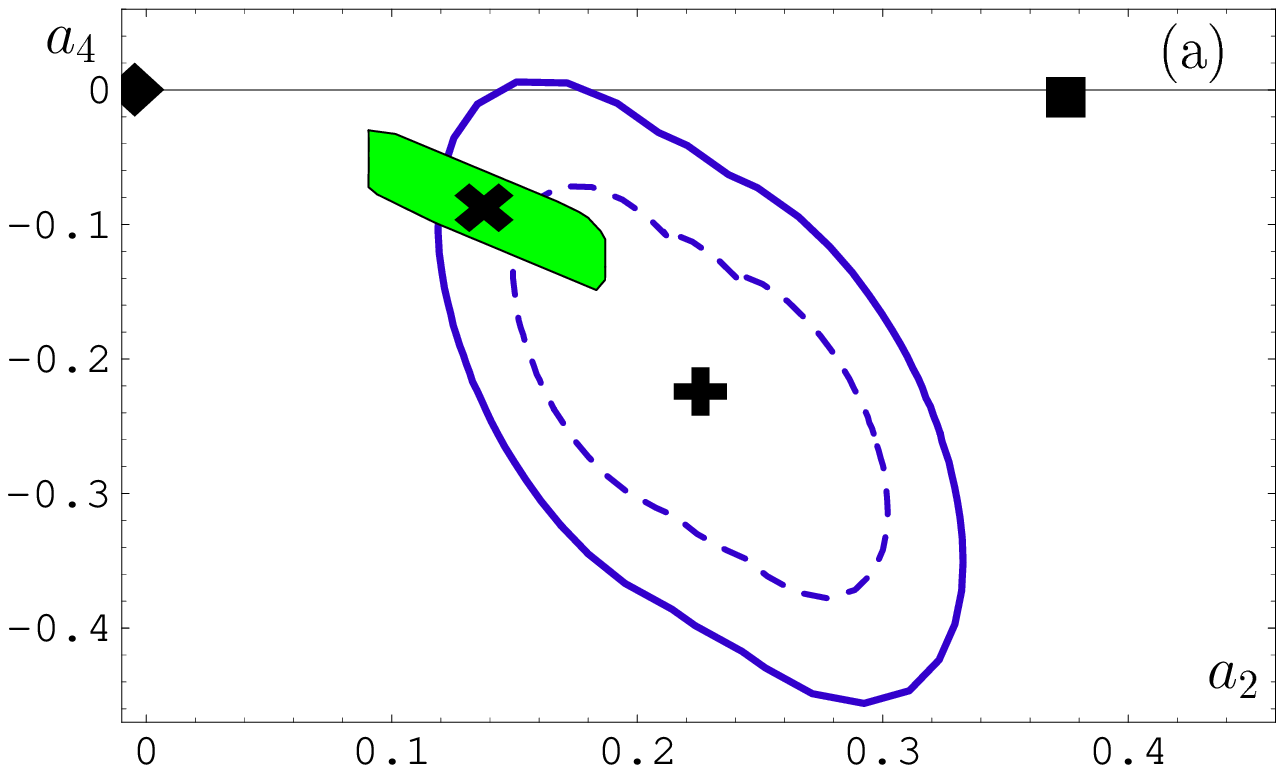}~~~
   \includegraphics[width=0.49\textwidth]{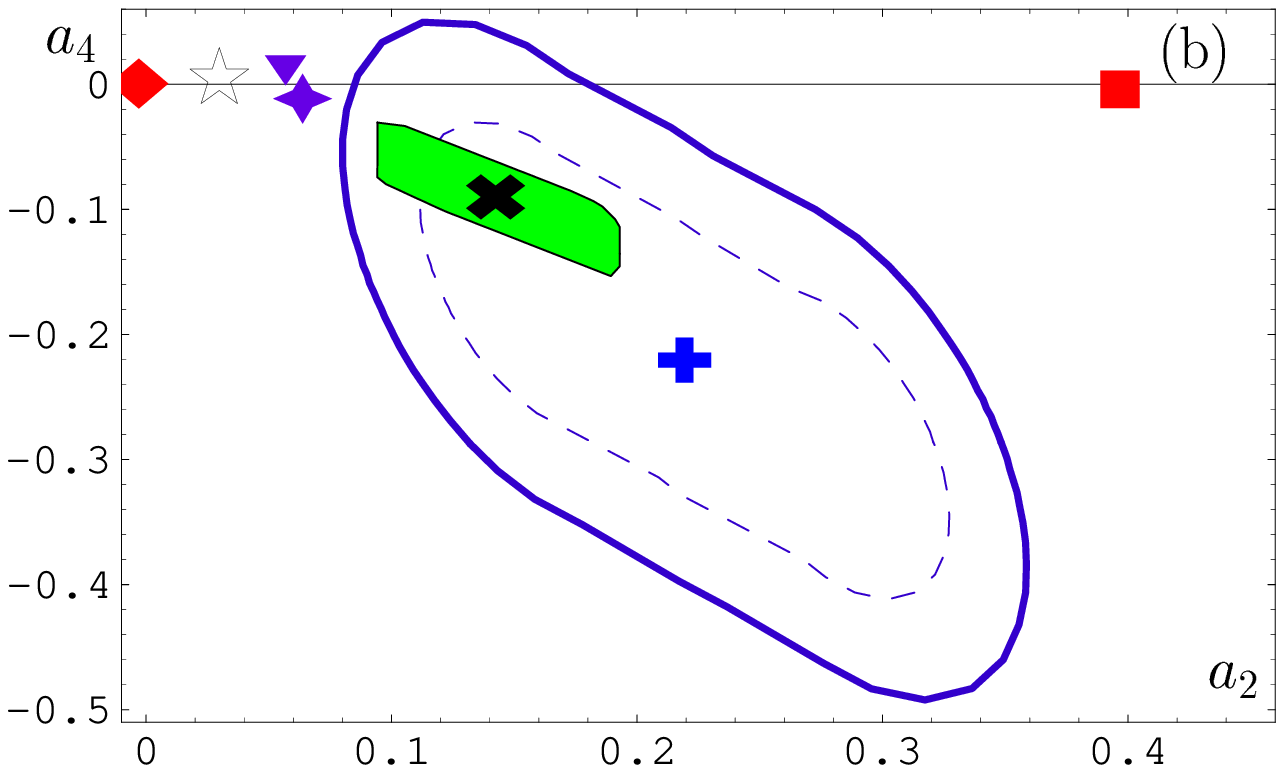}$$
   \vspace{-11mm}
\caption{Comparison of theoretical predictions of nonlocal QCD
         sum rules for $\lambda_q^2 = 0.4\text{ GeV}^2$
         and the CLEO data constraints obtained in the LCSR approach.
         \textbf{Left~(a):} Previous~\protect{\cite{BMS02}} results.
         \textbf{Right~(b):} New BMS~\protect{\cite{BMS03}} constraints.
         Here: {\ding{117}} = asymptotic DA,
         {\ding{54}} = BMS model, {\footnotesize\ding{110}} = {CZ} DA,
         {\ding{58}} = best-fit point, {\ding{73}~\protect{\cite{PPRWK99}}}
         and {\ding{70}~\protect{\cite{Pra01}}} =
         instanton models, {\footnotesize\ding{116}} =
         transverse lattice result~\protect{\cite{Dal02}}.
         All values are evaluated at $\mu^2_\text{SY}=(2.4~\text{GeV})^2$.}
\end{figure}

More recently~\cite{BMS03},
we have refined this extensive analysis
in several respects,
notably, by obtaining from the CLEO data
direct estimate for the inverse moment of the $\pi$DA
that plays a crucial role in pion electromagnetic/transition form factors
and by verifying the reliability
of the main results of the CLEO data analysis quantitatively.
We also refined our error analysis
by taking into account the total uncertainty of the twist-4 contribution
and treated the threshold effects in the strong running coupling more accurately.
The main upshot of this investigation is presented graphically in Fig.~4b,
where the values $\lambda_q^2 = 0.4\text{ GeV}^2$ and
$\delta_\text{Tw-4}^2 = 0.19(4)\text{ GeV}^2$ have been employed.
One can see that even with a 20\% uncertainty in the twist-4 contribution,
the CZ distribution amplitude ({\footnotesize\ding{110}}) is excluded
-- at least -- at the $4\sigma$-level,
while other well-known models (\ding{73}, \ding{70} and {\footnotesize\ding{116}})
with shapes more or less close to the asymptotic one (\ding{117})
are excluded  at the $2\sigma$-level.

These findings are further supported by extracting 
the inverse moment of the $\pi$DA 
from the CLEO data in a two-Gegenbauer model, 
$\displaystyle a_2+a_4=
 \langle x^{-1}\rangle^\text{exp}_{\pi}(\mu^2_0)/3-1$, 
at the low scale $\mu_0^2=1$ GeV$^2$. 
The obtained constraints are presented in Fig.~5b. 
One should compare them 
with the theoretical model-independent estimate 
of the inverse moment 
$\displaystyle \langle x^{-1}\rangle_{\pi}^\text{SR}(\mu^2_0 \approx 1~\text{GeV}^2)
 = 3.28 \pm 0.31$, 
obtained in the special NLC QCD sum rule 
using again $\lambda_q^2 = 0.4\text{ GeV}^2$~\cite{BM98,BMS01}, 
see Fig.~5a.
Noteworthily, these constraints match each other 
and both of them comply with the value 
$\frac{1}{3}\langle x^{-1} \rangle_{\pi} - 1 
 = 0.24 \pm 0.16$ 
found in~\cite{BK02} from a LCSR analysis 
of electromagnetic pion form factor. 
From Fig.~5b it is evident that again 
both the asymptotic $\pi$DA and the CZ model are far outside
the region of the CLEO experimental data.

\begin{figure}[t]%
$$\includegraphics[width=0.97\textwidth]{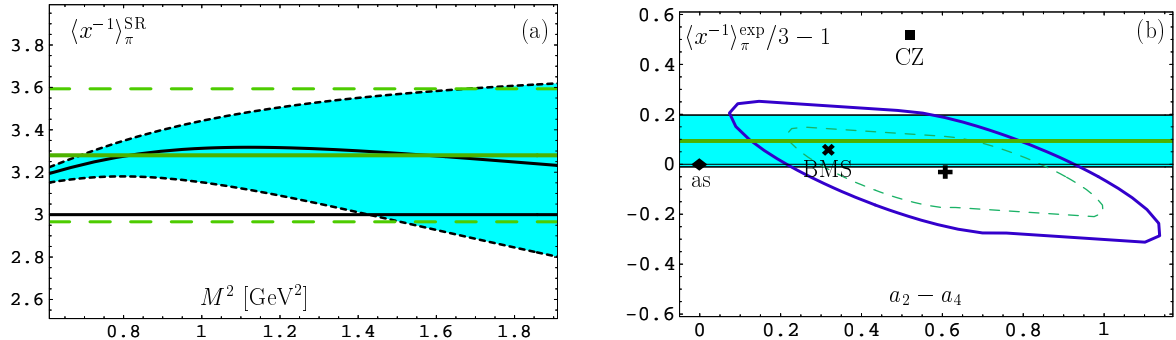}$$
   \vspace*{-11mm}
    \caption{\textbf{Left~(a):} The inverse moment 
        $\displaystyle \langle x^{-1} \rangle^\text{SR}_{\pi}$
        shown as a function of the Borel parameter $M^2$ from the
        special NLC QCD sum rule at the scale $\mu^2_0$ \protect\cite{BMS01};
        the light solid line is the estimate for
        $\displaystyle \langle x^{-1} \rangle^\text{SR}_{\pi}$;
        the dashed lines correspond to its error-bars.
    \textbf{Right~(b):} The result of the CLEO data processing for the quantity
        $\displaystyle \langle x^{-1} \rangle^\text{exp}_{\pi}/3-1$ at the scale
        $\mu^2_0 \approx 1~\text{GeV}^2$ in comparison with three
        theoretical models from QCD sum rules, CZ, BMS, and (a).
        The thick solid-line contour corresponds to the union of
       $2\sigma$-contours, while the thin dashed-line contour denotes
        the union of $1\sigma$-contours.
        The light solid line with the hatched band indicates the
        mean value of $\displaystyle \langle x^{-1} \rangle^\text{SR}_{\pi}/3-1$
        and its error bars in part (a).}
\end{figure}

\section{E791: Diffractive dijet production}
The Fermilab group E791 proposed~\cite{E79102} to exploit
experimentally the ideas on dijet diffractive dissociation
suggested in~\cite{FMS93} and further developed
in~\cite{NSS01,Che01,BISS02}. Braun et al.~\cite{BISS02} have used
a convolution-type approach to account for hard-gluon exchanges,
represented diagrammatically in the left part of Fig.~6.
\begin{figure}[hbt]
 $$\includegraphics[width=0.51\textwidth]{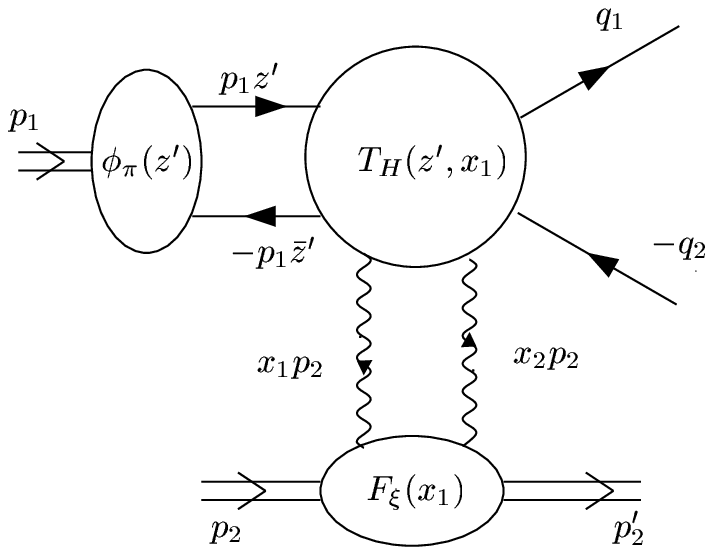}~~~
   \includegraphics[width=0.45\textwidth]{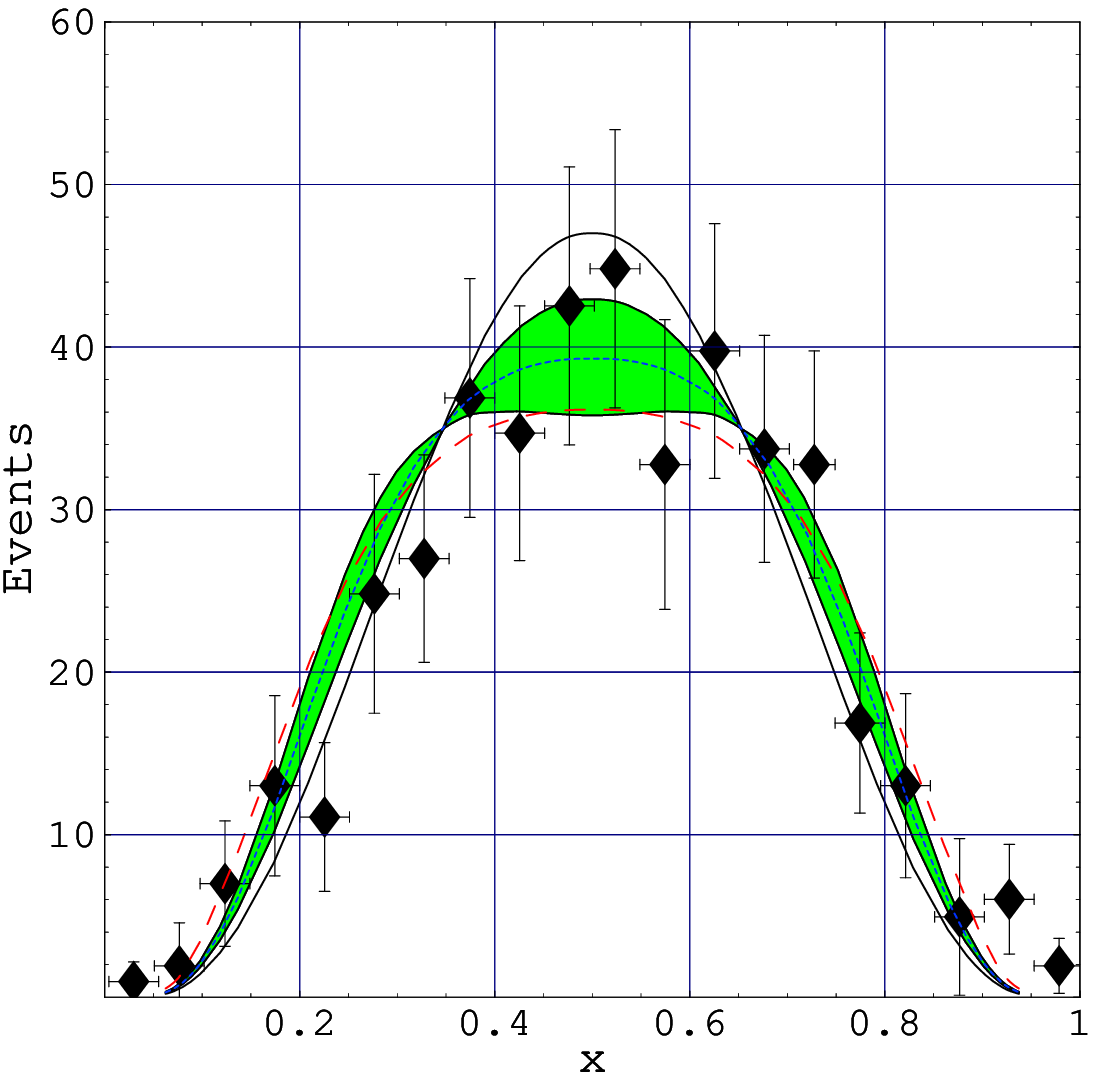}$$
   \caption{\textbf{Left:} Diffractive dijet $\pi A$-production in
   the E791 experiment with $q_{\perp}^2\simeq4\text{ GeV}^2$
   and $s\simeq1000\text{ GeV}^2$.
   \textbf{Right:} Asymptotic DA (solid line), CZ DA (dashed line)
   and  BMS bunch (shaded strip) in comparison with E791 data.
   Corresponding $\chi^2$ are: 12.56, 14.15 and 10.96
  (the last for BMS model with $\lambda_q^2=0.4\text{ GeV}^2$).}
\end{figure}
Following this convolution procedure (having also recourse to
\cite{FrMcD02}), and ignoring the distortion of our predictions
caused by the detector acceptance, we found the results displayed
in the right part of Fig.~6, making evident that, though the data
from E791 are not that sensitive as to exclude other shapes for
the pion DA (asymptotic and CZ model), also displayed for the sake
of comparison, they are relatively in good agreement with our
predictions. Especially, in the middle $x$ region, where our
$\pi$DAs -- the shaded strip -- has the largest uncertainties (see
Fig.~2a), the predictions are not in conflict with the data.
However, before this data set can be used for a quantitative
comparison, its inherent uncertainties have to be removed.

It is again worth emphasizing that because our model distribution
amplitudes -- exemplified by the BMS model --
are end-point suppressed (see Fig.~7),
they are not affected by the poor accuracy of the E791 experimental data
in these regions.
\begin{figure}[b]%
 \begin{minipage}{0.5\textwidth}
  $$\includegraphics[width=\textwidth]{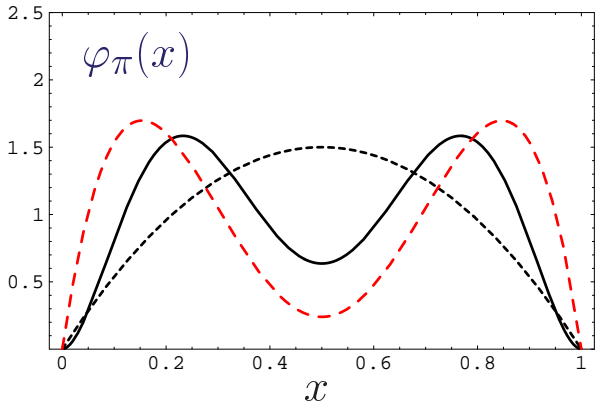}$$
 \end{minipage}~~
 \begin{minipage}{0.48\textwidth}
  \caption{Comparison of different DA curves
   aiming to illustrate the end-point suppression of the BMS model:
   CZ (dashed), asymptotic (dotted) and BMS (solid).}
 \end{minipage}
\end{figure}

\section{Conclusions}
Thanks to the recent high-precision CLEO experimental data~\cite{CLEO98}, 
we can answer more questions of nonperturbative QCD 
than a couple of years before. 
On the theoretical side, 
the method of QCD sum rules with nonlocal condensates~\cite{MR89,BM98,BM02,BMS01} 
provided a tool to determine more precisely 
than before a bunch of candidate DAs for the pion 
that are endpoint-suppressed 
due to a rather large QCD vacuum quark virtuality $\lambda_q^2$. 
On the other hand, 
the method of light-cone sum rules~\cite{Kho99,SchmYa99,BMS02} 
enables us to access the pion-photon transition form factor 
when one photon becomes real. 
Taking these theoretical approaches in conjunction, 
we were able to analyze the CLEO data at the NLO level 
in order to derive restrictive constraints 
on the first two Gegenbauer coefficients $a_2$ and $a_4$, 
which control the $x$-dependence
of the $\pi$DA. 
These parameters allow the reconstruction of the $\pi$DA 
and can be further tested against other experimental data, 
like those collected in the dijet production Fermilab experiment E791. 
The main conclusion 
is that both the CZ model as well as the asymptotic $\pi$DA 
are excluded---at least at the 2$\sigma$ level---by the CLEO data, 
while the two-humped end-point suppressed BMS distribution amplitude 
with a value of 
$\lambda_q^2\approx0.4~\text{GeV}^2$ 
is in a good agreement with the CLEO data 
and not in contradiction with the E791 data.
\bigskip

\textbf{Acknowledgments}\\
One of us (A.P.B.) would like to thank the organizers of the
Conference NAPP-2003 for the invitation and support.
This work was supported in part by INTAS-CALL 2000 N 587, the RFBR
(grant 03-03-16816), the Heisenberg--Landau Programme (grants 2002
and 2003), the COSY Forschungsprojekt J\"ulich/Bochum,
and the Deutsche Forschungsgemeinschaft.


\end{document}